\documentclass[review]{elsarticle} 

\usepackage{hyperref}
\usepackage{lineno}

\usepackage{siunitx}

\usepackage{tabularx}

\usepackage{listings}

\usepackage{subcaption}

\usepackage{xcolor}
\lstdefinestyle{mystyle}{
	basicstyle=\scriptsize\ttfamily,
	breakatwhitespace=false,         
	breaklines=true,                 
	captionpos=b,                    
	keepspaces=true,                                   
	showspaces=false,                
	showstringspaces=false,
	showtabs=false,                  
	tabsize=2,
	frame = single, 
	backgroundcolor = \color[RGB]{255,255,235}
}

\usepackage[T1]{fontenc}

\journal{Nuclear Instruments and Methods in Physics Research Section A: Accelerators, Spectrometers, Detectors and Associated Equipment}

\begin{document}

\begin{frontmatter}

\title{megas: development and validation of a new simulation tool for the Micromegas detectors.}

\author[address-ub,address-ifin]{Dan Andrei Ciubotaru\corref{mycorrespondingauthor}}
\cortext[mycorrespondingauthor]{Corresponding author}
\ead{dan.andrei.ciubotaru@cern.ch}
\author[address-ifin]{Michele Renda}

\address[address-ub]{Faculty of Physics, University of Bucharest, Bucharest - M\u agurele, Romania }
\address[address-ifin]{Departament of Elementary Particle Physics, IFIN-HH, Reactorului 30, RO-077125, P.O.B. MG-6, M\u agurele, Romania}

\begin{abstract}
We present \texttt{megas}, a new software tool that improves the simulation of the Micromegas gas detectors. Our tool offers the possibility to configure multiple arrangements with one or more layers of MM detectors. A series of simple commands can easily modify the constructive properties of each of these detectors, such as dimensions, gas composition, and electric field. \texttt{megas} is based on \texttt{Betaboltz} \cite{renda2020betaboltz, betaboltz_git}, an open-source library that simulates the step by step movement of electrons in an electric field using Monte-Carlo methods.
A real-life example was simulated, and to validate our application, we perform a detailed analysis of our data compared with the actual experimental results. We present a selection of data obtained simulating real user-case scenarios, compared with results available in the literature, especially in the upgrade phase of the Muon spectrometer of the ATLAS detector at LHC.
\end{abstract}


\end{frontmatter}


\section{Introduction}
Detectors using  Micromegas (MM) technology represents on of the best options in experimental physics when considering the detection of ionizing particles. This reputation was built in the past years based on the fast signal generation  ($<1$ \si{\nano\second}) at high particle rates and good spatial resolution with values under 100 \si{\micro\meter} \cite{giomataris}. This detector technology consists of two parallel sectors separated by a micromesh and limited at both ends by a cathode and an anode. The first region, commonly about \SI{5}{\milli\meter}, is called the drift region and the second one, considerably smaller, about \SI{128}{\micro\meter}, is named avalanche region. The voltage applied to these electrodes has the role of creating a conversion space in the drift region and a multiplication effect in the avalanche region, generating the signal. These unique aspects, together with the relatively low cost per sensitive area, qualified this technology as the best option for the New Small Wheel \cite{AtlasTDR20} detector of the ATLAS experiment at the Large Hadron Collider (LHC) \cite{mmInATLAS}.

The increased use  of this detector required the development of specific computer simulation software that can be used to facilitate improvements and optimizations. These should be efficient in terms of resources and development time. While there are tools that allow the simulation of the gas detectors, at this time, none of them are suitable for the analysis presented in this article.

\texttt{Geant4} \cite{geant4} may be considered the first choice when simulating the passage of a particle through matter, but it is unable to perform microscopic simulations of electrons in gaseous medium, essential for performance studies (resolution, response time, etc.) of gas-filled detectors.

\texttt{Garfield} \cite{garfield} and a most recent version of it, \texttt{Garfield++} \cite{garfield++}, are some of the tools utilized to simulate gas detectors in high energy physics. \texttt{Garfield} uses an analytical approach when determining the electrical field, and there is an integration with \texttt{MagBoltz} \cite{biagi}\cite{magboltz} to calculate the Townsend coefficient. This configuration comes with several limitations:
\begin{itemize}
	\item the electrical field is determined analytically and can only be uniform; \footnote{a possible integration with a three-dimensional filed solver \texttt{neBEM} \cite{nebem} is considered \cite{schindler}}
	\item the simulated gas mixtures are limited and are hard-coded into \texttt{Magboltz};
	\item consist in a mixture of programming languages in particular C++ and Fortran, with the latter being quite difficult to maintain;
	\item is unable to run on multi-core, making it difficult to take advantage of modern CPU architectures.
\end{itemize}
	
Given these shortcomings, we designed an application focused on simulating Micromegas detectors, highly customizable through specific parameters (see table \ref{tab:parameters}).

This application is written in C++ 17, runs on multi-core and integrates the \texttt{Betaboltz} library \cite{renda2020betaboltz, betaboltz_git} to make use of a microscopic simulation approach. 

Since \texttt{Betaboltz} reads the cross-section tables required to reproduce the interaction processes from \texttt{LXCat} \cite{lxcat}, the number of gases is limited only to the availability of the data in the database or literature. 

Currently, the electrical field is considered uniform, but any field can be expressed analytically (through a specialization of the class \texttt{BaseField}).

This article is organized as follows. Section \ref{sec:setup-implementation} describes how the application was implemented, along with how we configured the simulation setup: from the core components of the application, the experimental setup used along with the input parameters, to how the spatial resolution and drift speed were determined.

Section \ref{sec:results} is reserved for the presentation of the results obtained when modifying some of the detector constructive parameters: the electrical field of the drift region, the angle of the incident particle, the gas mixture used, and the number of primary particles used.

In section \ref{sec:conslusions}, we present the conclusions obtained from the development and validation of this application.

\section{Simulation setup and implementation}
\label{sec:setup-implementation}
In this section, we describe the \texttt{megas} application and the tools used to obtain the results presented in the next section.

\subsection{Program summary}
\begin{figure}[h]
	\includegraphics[width=.4\columnwidth]{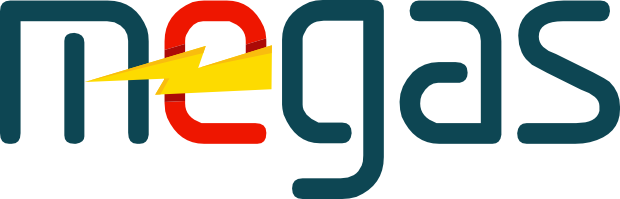}
	\centering
	\label{fig:megas-logo}
\end{figure}
\emph{Program title:} \texttt{megas}

\emph{Licensing provisions:} LGPL v3

\emph{Programming language:} C++17

\emph{URL:} https://gitlab.com/dan.ciubotaru/megas.git \cite{megas_git}

\subsection{Simulation framework}
\begin{figure}[ht]
	\includegraphics[width=.85\columnwidth]{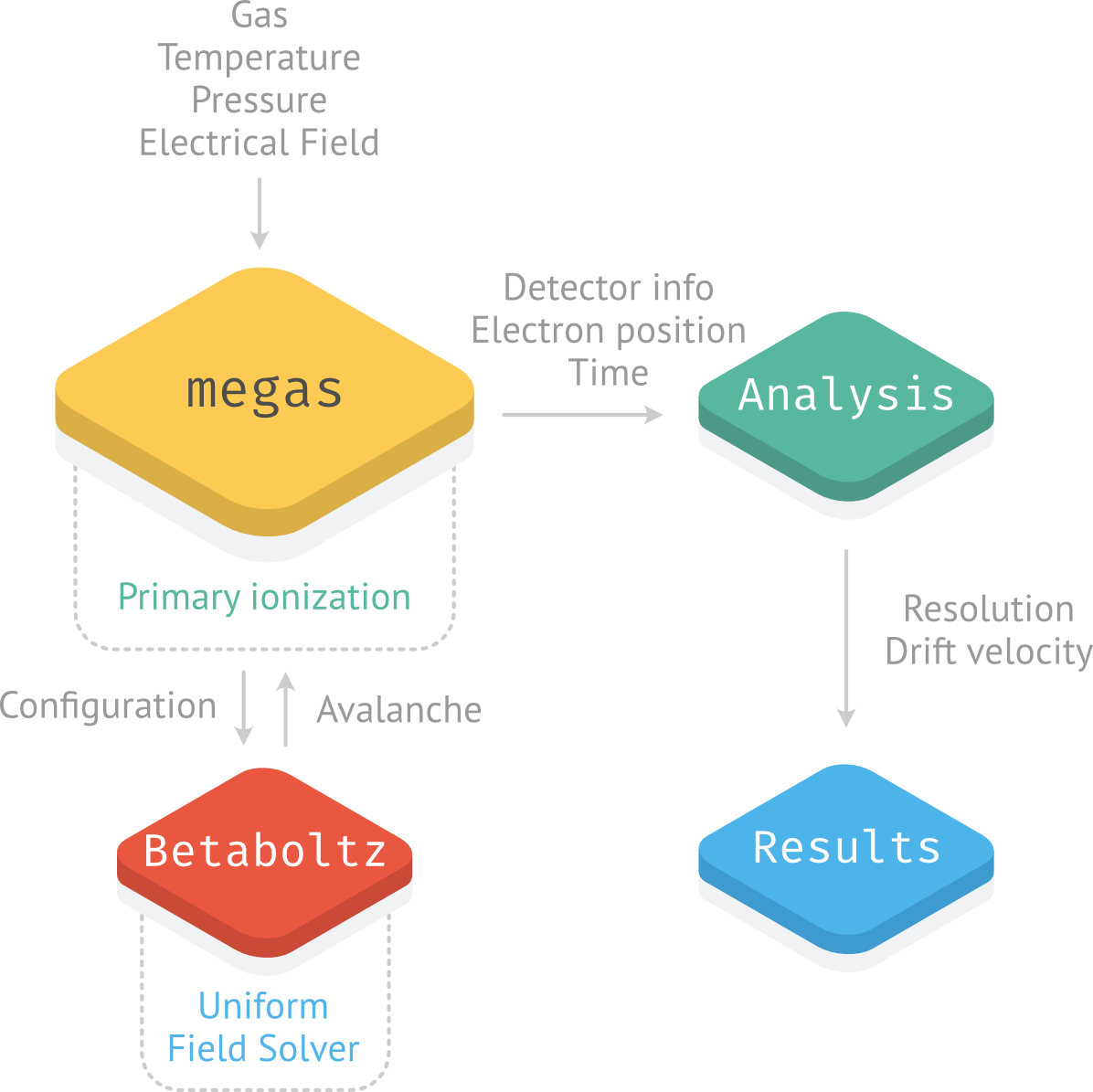}
	\centering
	\caption{\texttt{megas} simulation flow}
	\label{fig:megas_sim_chain}
\end{figure}

The new proposed tool is developed in C++ and relies on the \texttt{Betaboltz} library to simulate the movement of electrons and ions in the sensitive area of the Micromegas detector. Figure \ref{fig:megas_sim_chain} outlines the \texttt{megas} workflow from input parameters to analysis results.

The core application functionality relies on three classes: 
\begin{itemize}
	\item \texttt{MicromegasDetector}; 
	\item \texttt{MicromegasSetup};
	\item  \texttt{SimulationLimiter}.
\end{itemize}
This design extends the abstract classes available in the \texttt{Betaboltz} framework. This approach allowed us to have only the code specific for Micromegas simulation, delegating to \texttt{Betaboltz} the code common to general gas detector simulation, making \texttt{megas} more maintainable and compact.

\begin{figure}[h]
	\includegraphics[width=.82\columnwidth]{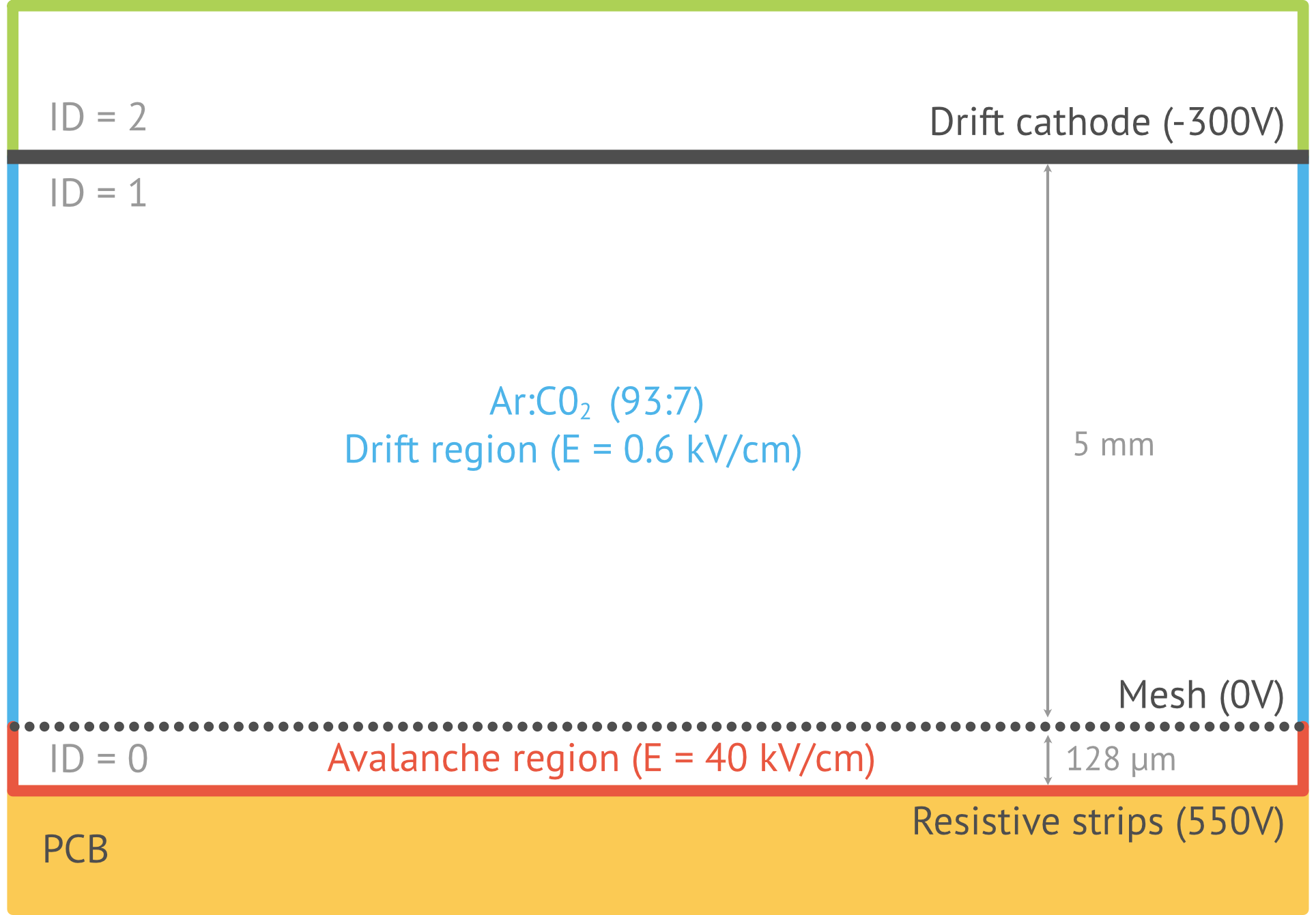}
	\centering
	\caption{Schematic representation of the Micromegas detector and the default simulation parameters.}
	\label{fig:detector-description}
\end{figure}

\texttt{MicromegasDetector}: within this class all the technical details of the MM detector are described, including dimensions, electric field, and gas mixture. Although more information about the geometry is related in the next section, it is worth mentioning that our application uses volumes  (see fig. \ref{fig:detector-description}) to describe the regions of the MM detector: avalanche, drift, and the 'upper' section \footnote{we use this term to express the area above the cathode}. In this way, we created the \texttt{getVolumeIds()} method to know precisely when the electron changes a certain region or leaves the sensitive area, and simulation needs to be halted.

Starting from the concept of volumes presented above, through the \texttt{BaseField} method, the application is designed to be able to change the electric field separately for each volume by controlling the mesh, cathode, and anode voltage.

\texttt{MicromegasSetup}: through this class, we can configure one or multiple layers of MM detectors in different arrangements and orientations, the particle gun angle, and the number of the primary interactions in each detector.

\texttt{SimulationLimiter} is overriding the \texttt{isOver()} method from \texttt{BaseBulletLimiter} class, inherited from \texttt{Betaboltz} and it makes possible to halt the simulation when all electrons leave the drift region - the sensitive area of the simulated detector.

The number of primary electrons ($N_p$) generated by the incident particles, in our case, muons, remains an essential factor in our application. To estimate this value, we relied on a simple composition law for gas mixtures described by Claus Grupen and Irene Buvat \cite{GupenBuvat}. For the chosen height of \SI{5}{\milli\meter} of the drift region, we obtained a value of 50 primary electrons generated in this region, distributed uniformly and random across the muon path.

\subsection{Geometry and layout}
\begin{figure}[ht]
	\includegraphics[width=.82\columnwidth]{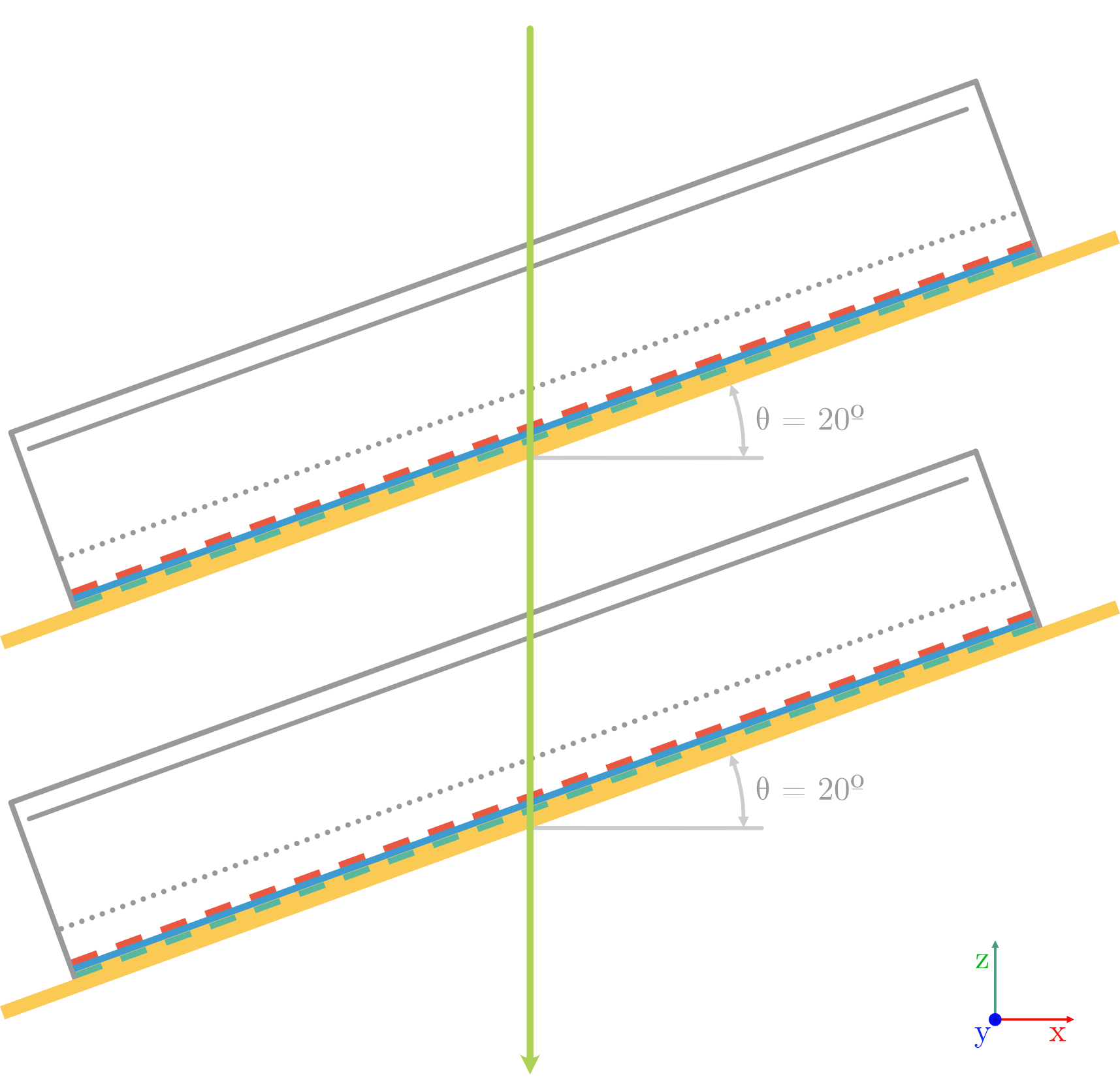}
	\centering
	\caption{Example of a simulation run for a $\theta = 20$ degree angle. The green line represents the virtual trajectory of the muons.}
	\label{fig:detector-layout}
\end{figure}

Our tool offers the possibility to configure multiple setups with one or more layers of MM detectors. For this study, we choose to replicate a real-life scenario, presented in ATLAS New Small Wheel Technical Design Report \cite{AtlasTDR20}. The setup consists of two Micromegas detectors 20 x 20 \si{\centi\meter}, stacked one atop the other, with the same orientation (see fig. \ref{fig:detector-layout}) and a distance of \SI{1}{\meter} between  them. The gap separating the detector layers is void and does not affect the simulation results. The height of the amplification space is \SI{128}{\micro\meter}, and the height of the drift region is \SI{5}{\milli\meter}.

The reference gas mixture is $Ar:CO_2$ in a proportion of 93:7 at normal temperature and pressure conditions (NTP), \SI{20}{\degreeCelsius} and \SI{101.325}{\kilo\pascal}. The scattering cross sections dataset for Ar was taken from Biagi \cite{biagi2019lxcat} and for $CO_2$ from Phelps \cite{bulos_excitation_1976, phelps2019lxcat, phelps}.
Using the voltage of  \SI{-300}{\volt} for the cathode, \SI{0}{\volt} on the mesh, and \SI{550}{\volt} on the anode, we induce an electrical field of \SI{40}{\kilo\volt\per\centi\meter} in the amplification region and \SI{0.6}{\kilo\volt\per\centi\meter} on the drift region, in the normal direction to the detector plane.

\subsection{Simulation configuration and parameters}
While the default parameters are listed in the previous section, we implemented the possibility to change them through arguments passed to the main program (see table \ref{tab:parameters}). This feature offers the possibility to test different simulation scenarios easily.

\begin{table*}
	\begin{tabularx}{\textwidth}{  l  l X }
		\hline
		\textbf{Parameter} & \textbf{Command}              & \textbf{Description}                                                      \\ \hline
		run                & \texttt{-r, -{}-run}          & ID for a specific simulation                                              \\
		events             & \texttt{-e, -{}-events}       & Number of events to simulate                                              \\
		threads            & \texttt{-j, -{}-jobs}         & Number of concurrent threads                                              \\
		angle              & \texttt{-a, -{}-angle}        & Angle of the incident particles [\si{\deg}]                               \\ 
		interaction        & \texttt{-i, -{}-interactions} & Primary electrons $N_p$ generated per unit of length [\si{\milli\meter}]  \\
		cathode voltage    & \texttt{-{}-cathode-voltage}  & Cathode voltage [\si{\volt}]                                              \\
		mesh voltage       & \texttt{-{}-mesh-voltage}     & Mesh voltage [\si{\volt}]                                                 \\
		anode voltage      & \texttt{-{}-anode-voltage}    & Anode voltage [\si{\volt}]                                                \\
		gas or gas mixture & \texttt{-g, -{}-gases}        & Gas mixture composition.\newline(e.g. \texttt{Ar:93:Biagi,CO2:7:Phelps})  \\
		gas pressure       & \texttt{-p, -{}-pressure}    &  Gas pressure  [\si{\kilo\pascal}]                                         \\
		\hline
	\end{tabularx}
	\caption{Parameters of the \texttt{megas} application}
	\label{tab:parameters}
\end{table*}

\subsection{Determining the spatial resolution and the longitudinal drift velocity}\label{determining-resolution&velocity}
\subsubsection{Spatial resolution}
\begin{figure}[h]
	\includegraphics[width=.85\columnwidth]{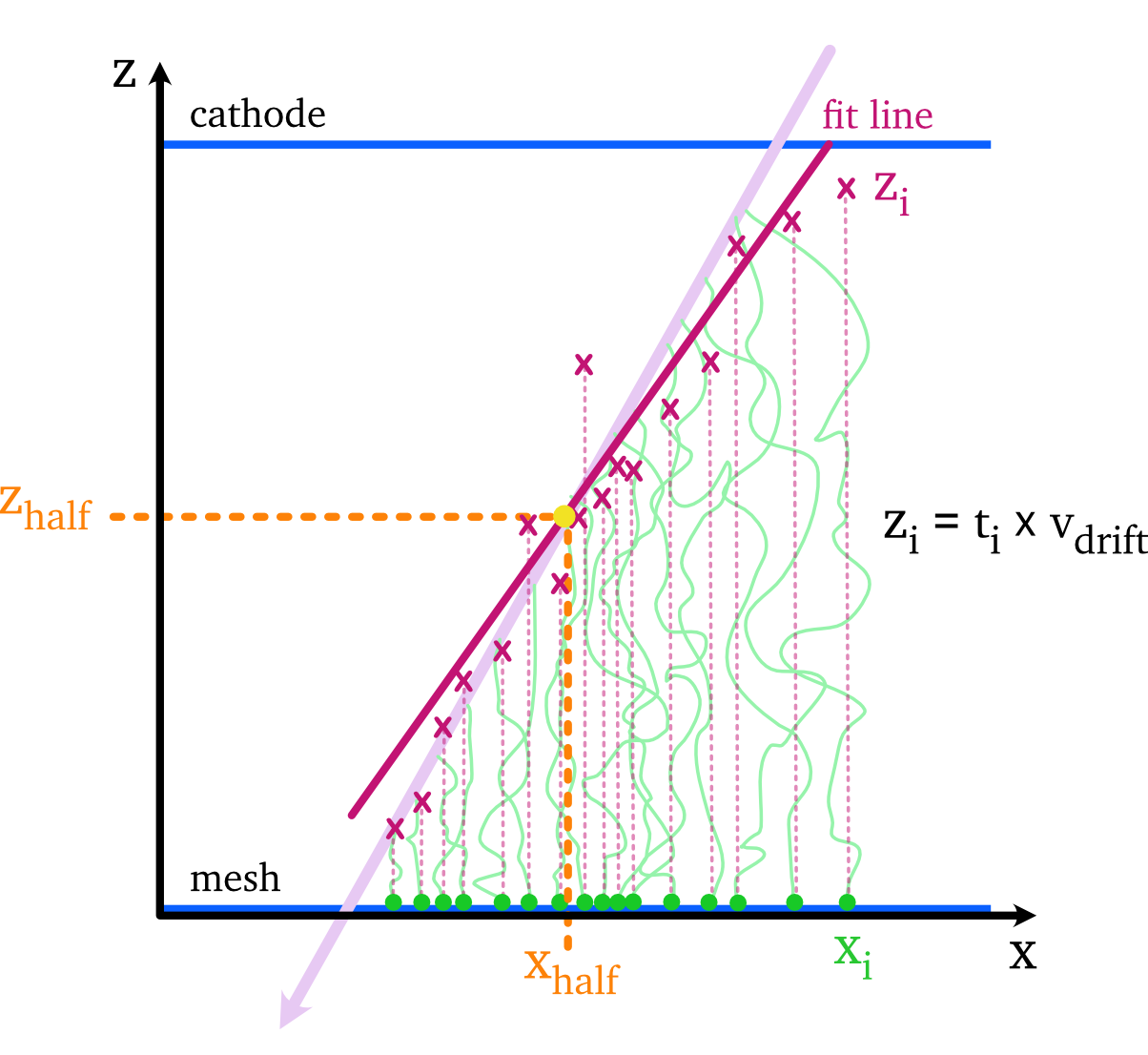}
	\centering
	\caption{Schematic representation of the $\mu$TPC method. The arrow represents the muon path crossing the sensitive area, the green lines the drift of the primary electrons, the X the reconstructed points used to determinate the fit line. }
	\label{fig:uTPC-draw}
\end{figure}

To calculate the spatial resolution, we used the $\mu$TPC method \cite{ALEXOPOULOS2010161, AtlasTDR20}. This procedure estimates the coordinate of the primary ionization $z_i$ for each electron $i$ as $z_i = t_i \times  v_{drift}$. Knowing the drift velocity $v_{drift}$ in $Ar:CO_2$ (93:7) as \SI{4.7}{\centi\meter\per\micro\second} \cite{iodice}, time ($t_i = t_i^{finish} - t_i^{start}$), and the position $x_i$ where the electron leaves the drift volume, we are able to compute $z_i$. Identifying all $z_i$ positions we can reconstruct a local track by fitting with a straight line (see fig. \ref{fig:uTPC-draw}). The hit position, $x_{half}$, is determined for each detector as the intersection of the fitted line with the half-height plane of the drift region, in our case at \SI{2.5}{\centi\meter}. The distribution of the difference between the two $x_{half}$ values is fitted adequately by a gaussian, and the spatial resolution for one chamber is calculated as the standard deviation of this fit divided by $\sqrt{2}$ (std is computed using the \texttt{norm.fit()} method from \texttt{scipy.stats} \texttt{Pyhton} package).

We want to point out that the spatial resolution calculated with our application is an intrinsic one and is specific to the gas mixture used for the given drift height. Consequently, the calculated intrinsic resolution is not influenced by the mesh transparency, strips width, and strips gap.

\subsubsection{Drift velocity}
Similarly, with the calculation of spatial resolution, we rely on the same initial and final values of the time and position for each primary interaction to determine the electron path. The distance vector is computed and we obtain the component along the z-axis, normal to the electrical field. The longitudinal drift velocity is expressed as the distance divided by time, averaged for all interactions in a given simulation.

\subsection{Limitations of  \texttt{megas}}
Regardless of how accurately one tries to model real-life Micromegas detector, there are always some drawbacks that come with a simulation software. In particular, for \texttt{megas}, we identified a few limitations:
\begin{itemize}
	\item the primary generated electrons (10/mm) are distributed randomly and evenly along the muon path;
	\item delta ray that can potentially cause secondary ionizations are not considered;
	\item the simulation does not take into account the pillars and mesh transparency;
	\item to reduce the computing time, the avalanche is not simulated, and the electron path is stopped when it leaves the drift region;
	\item we decided not to calculate the signal induced on the readout strips.
\end{itemize}

\section{Results}
\label{sec:results}

\begin{figure*}
	\centering
	\begin{subfigure}{.48\textwidth}
		\centering
		\includegraphics[width=\linewidth]{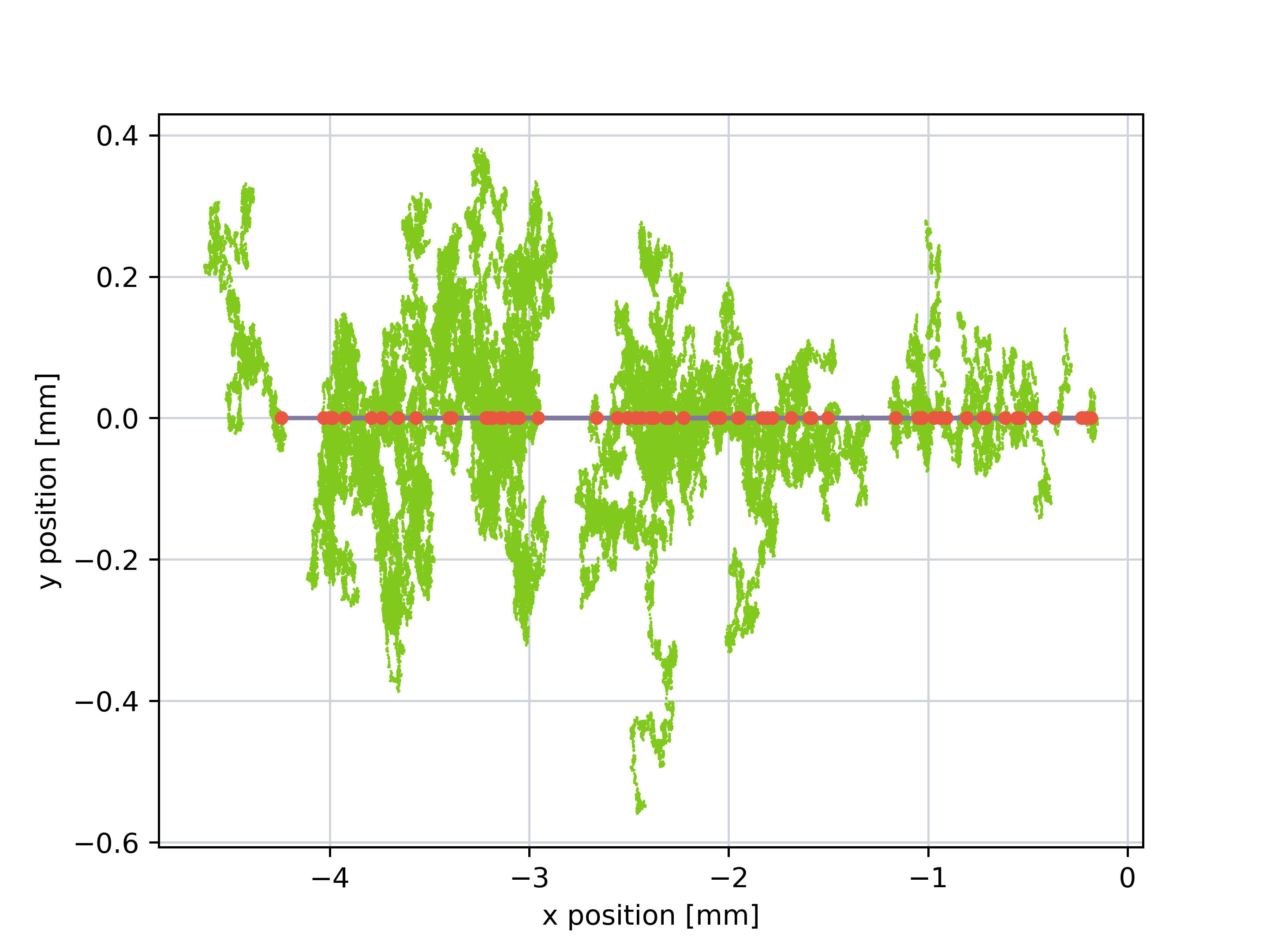}
		\caption{}
		\label{fig:sim_angle_primary_xy}
	\end{subfigure}
	\begin{subfigure}{.48\textwidth}
		\centering
		\includegraphics[width=\linewidth]{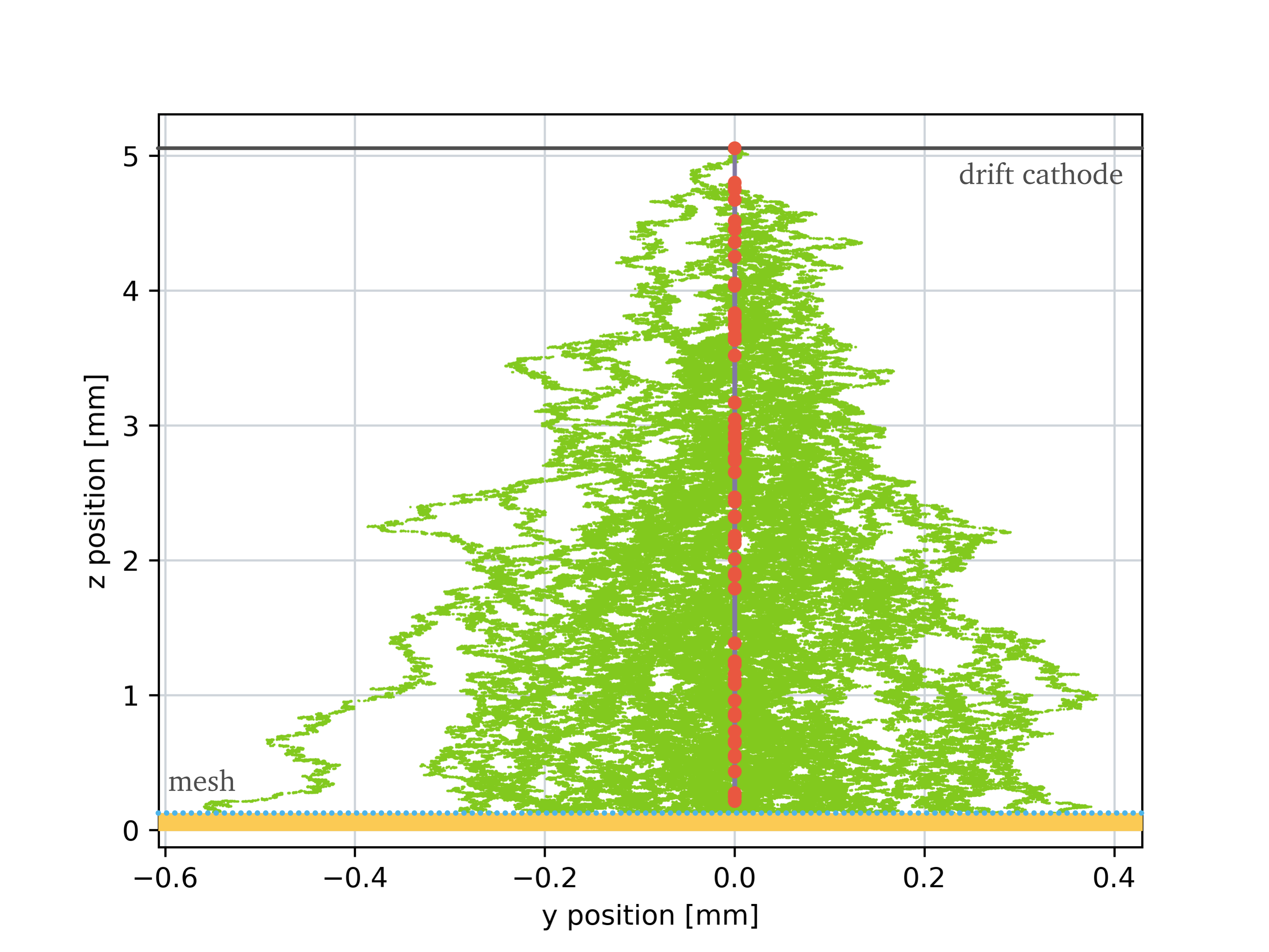}
		\caption{}
		\label{fig:sim_angle_primary_yz}
	\end{subfigure}
	\begin{subfigure}{.48\textwidth}
		\centering
		\includegraphics[width=\linewidth]{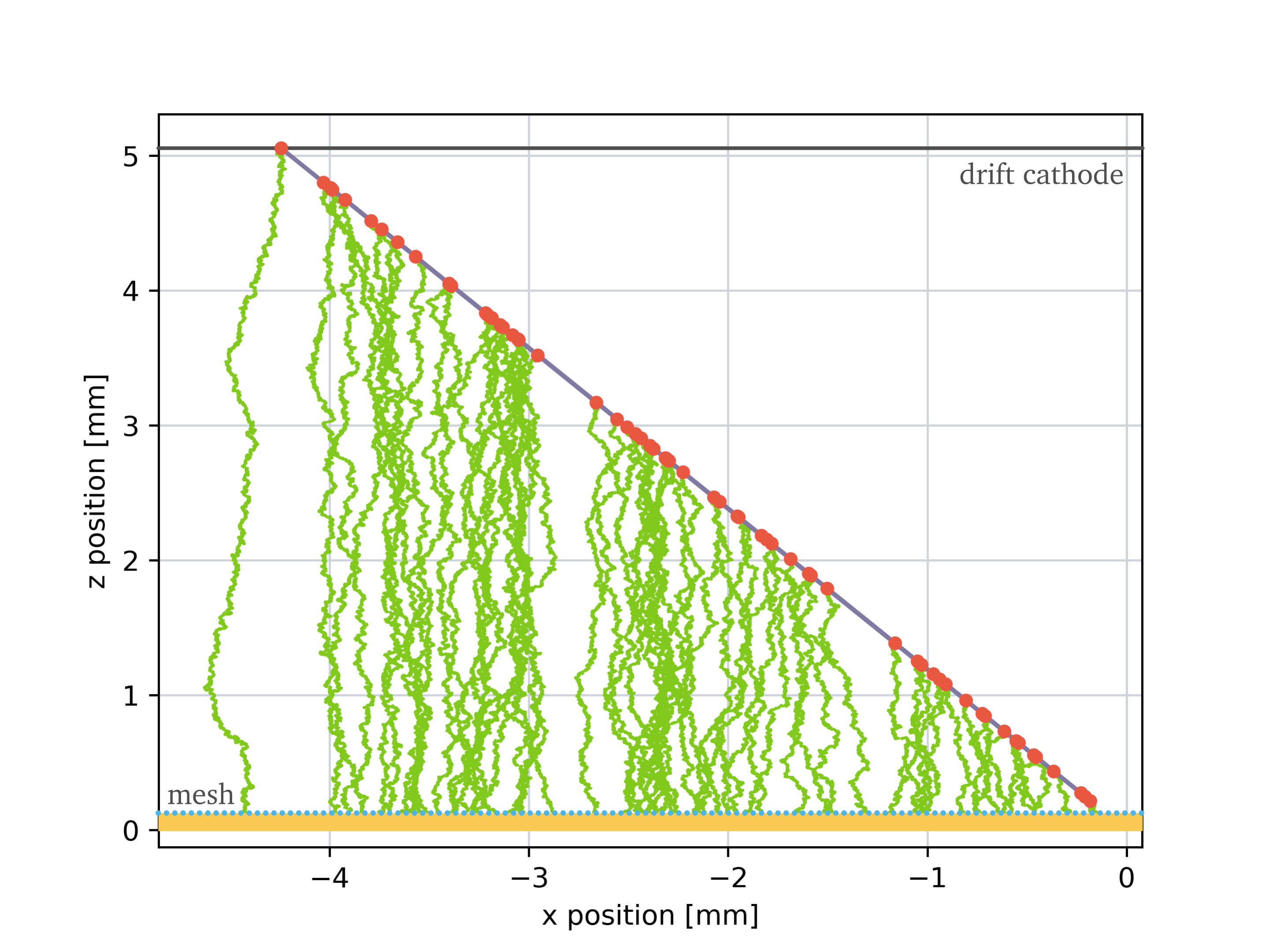}
		\caption{}
		\label{fig:sim_angle_primary_xz}
	\end{subfigure}
	\caption{Simulation of the electron drift path for a single event of a crossing muon with an angle $\theta=\SI{40}{\deg}$. The muon produced \SI{9.7}{pairs\per\centi\meter}, for a total of \SI{63}{} primary ionizations, which are drifted by an electric field of \SI{0.6}{\kilo\volt\per\centi\meter} aligned along the detector axis. A total of \SI{691460}{} collisions where simulated.  Gas mixture $Ar:CO_2$ $93:7$  at \SI{101.325}{\kilo\pascal} with  cross section tables respectively from Biagi \cite{biagi2019lxcat} and Phelps\cite{phelps2019lxcat}.}
	\label{fig:sim_angle_primary}
\end{figure*}

To test and validate the application, several simulation scenarios were considered. This was done by modifying the essential parameters, focusing on two particular properties of the Miromegas detector: the drift velocity along the electrical field and the spatial resolution (see section \ref{determining-resolution&velocity}). In fig. \ref{fig:sim_angle_primary_xy} it is possible to visualize the path of the primary electrons generated by an incident muon producing \SI{9.7}{pair\per\milli\meter} withing the sensitive area. It is possible to see how the transversal drift of the electrons directly affects the reconstruction of the muon path, limiting the intrinsic detector resolution. While single event simulation may be useful to get qualitative knowledge of the processes taking place within the detectors' sensitive region, a quantitative description can only be done if we perform an average over multiple events. In the next subsection are presented some different simulation setups where we modified one detector parameter. We analyzed the different response averaged over a high number of collision events.

\subsection{Simulation of  the voltage changes on the cathode}
\begin{lstlisting}
megas --angle 20 --events 3000 --interactions 10 --cathode-voltage $voltage
\end{lstlisting}
One of the most critical parameters of the MM detector construction is the voltage applied between the cathode and the mesh, in the drift region.  For this simulation, we maintain all the application parameters to the default values, except the number of simulated events and the voltage applied to the drift cathode.
\begin{figure}
	\includegraphics[width=.82\columnwidth]{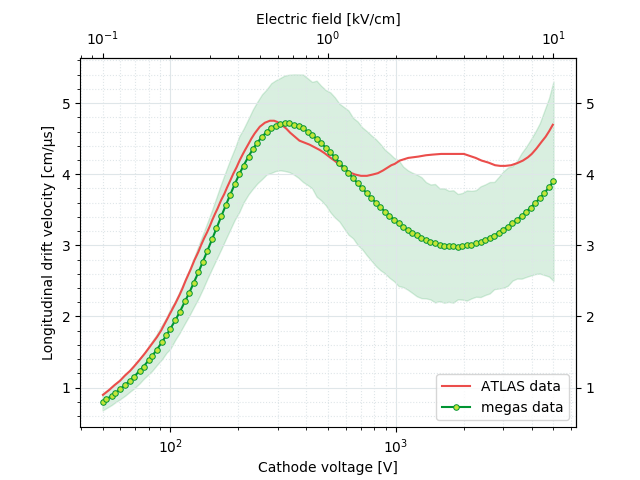}
	\centering
	\caption{Longitudinal Drift velocity for $V_{cathode}$ between \SI{50}{\volt} and \SI{5}{\kilo\volt} ($E_{drift}$ between \SI{0.1}{\kilo\volt\per\centi\meter} and \SI{10}{\kilo\volt\per\centi\metre}, \SI[per-mode=symbol]{10}{\per\milli\meter} primary electrons, and beam angle of \SI{20}{\deg}. Simulation on \SI{3000}{} events in $Ar:CO_2$ $93:7$ gas mixture at \SI{101.325}{\kilo\pascal}.  Cross section tables respectively from Biagi \cite{biagi2019lxcat} and Phelps\cite{phelps2019lxcat} with reference values digitized from ATLAS New Small Wheel Technical Design Report \cite[, fig. 5.12]{AtlasTDR20}. The green band shows the standard deviation of the drift velocity over \SI{3000}{} events.}
	\label{fig:sim_voltage_velocity}
\end{figure}
\begin{figure}
	\includegraphics[width=.82\columnwidth]{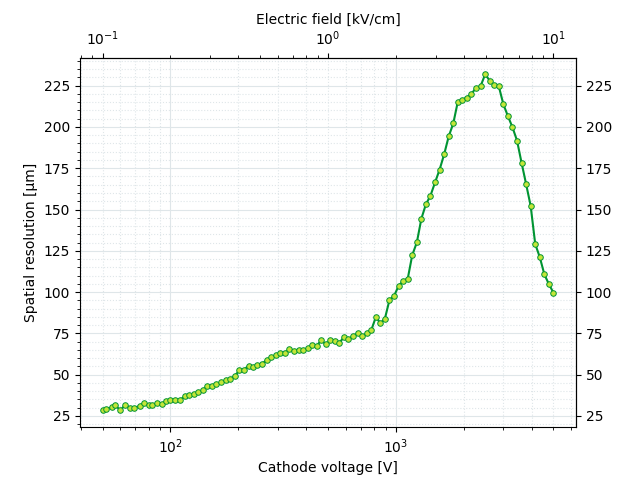}
	\centering
	\caption{Spatial resolution for $V_{cathode}$ between \SI{50}{\volt} and \SI{5}{\kilo\volt} ($E_{drift}$ between \SI{0.1}{\kilo\volt\per\centi\meter} and \SI{10}{\kilo\volt\per\centi\metre}, \SI[per-mode=symbol]{10}{\per\milli\meter} primary electrons, and beam angle of \SI{20}{\deg}. Simulation on \SI{3000}{} events in $Ar:CO_2$ $93:7$ gas mixture at \SI{101.325}{\kilo\pascal}.  Cross section tables respectively from Biagi \cite{biagi2019lxcat} and Phelps\cite{phelps2019lxcat}.}
	\label{fig:sim_voltage_resolution}
\end{figure}

\begin{figure}
	\includegraphics[width=.82\columnwidth]{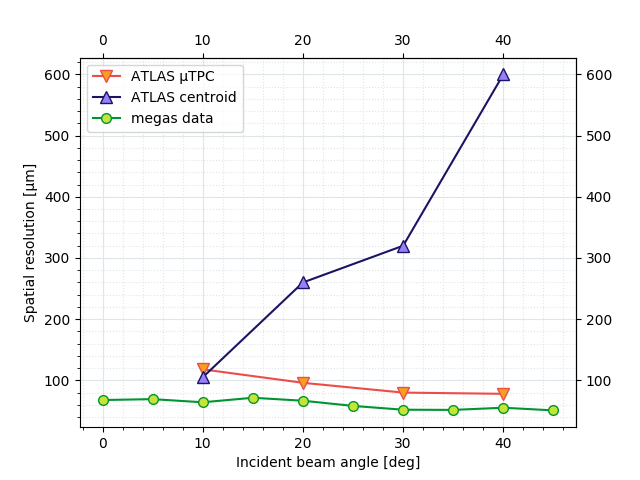}
	\centering
	\caption{Spatial resolution for different beam angles between \SI{0}{\deg} and \SI{45}{\deg}. We calculated the results for the spatial resolution using the $\mu$TPC method. Simulation on \SI{3000}{} events in $Ar:CO_2$ $93:7$ gas mixture at \SI{101.325}{\kilo\pascal}.  Cross section tables respectively from Biagi \cite{biagi2019lxcat} and Phelps\cite{phelps2019lxcat} with reference values digitized from ATLAS New Small Wheel Technical Design Report \cite[, fig. 5.11]{AtlasTDR20}}
	\label{fig:sim_angle_resolution}
\end{figure}

We can observe that for voltages below 1000 V, the results obtained for the longitudinal drift velocity (see fig. \ref{fig:sim_voltage_velocity}) are in accordance with ATLAS data\cite{AtlasTDR20}. When talking about the spatial resolution (fig. \ref{fig:sim_voltage_resolution}), for the same threshold of \SI{1000 }{\volt} the obtaining values fall under the limit of \SI{100}{\micro\meter} established in the upgrade specifications for the New Small Wheel \cite{AtlasTDR20}.

\bigskip
\subsection{Simulation of  the incident particle angle}
\begin{lstlisting}
megas --angle $angle --events 3000 --interactions 10
\end{lstlisting}
One important application of the Micromegas detector is for the New Small Wheel sector of the ATLAS detector, tracking muons coming at different angles in the pseudorapidity interval of $1.0 < |\eta| < 2.7$, relative to the beam axis. In this situation, it is important to study the spatial resolution and the drift velocity when changing the angle of the incident particles.

We notice that when changing the incident angle (see fig. \ref{fig:sim_angle_resolution}), there is a decrease in the spatial resolution from approx. \SI{70}{\micro\meter} at \SI{0}{\deg} to approx. \SI{50}{\micro\meter} at \SI{45}{\deg}. This drop in resolution is due to the increasing number of primary interactions generated by the incident particle traversing a longer distance in the drift region. Compared with ATLAS $\mu$TPC data, the spatial resolution achieved is below \SI{100}{\micro\meter} following ATLAS New Small Wheel Technical Design Report \cite{AtlasTDR20} upgrade specification.

\subsection{Simulation of different gas mixtures}
\begin{lstlisting}
megas --angle 20 --events 5000  \
--gases "Ar:$((ratio)):Biagi,CO2:$((100-ratio)):Biagi"

megas --angle 20 --events 5000  \
--gases "Ar:$((ratio)):Biagi,CO2:$((100-ratio)):Bordage"

megas --angle 20 --events 5000  \
--gases "Ar:$((ratio)):Biagi,CO2:$((100-ratio)):Hayashi"
\end{lstlisting}
The possibility of modifying the gas mixture is a significant advantage that \texttt{megas} offers, gas mixtures representing the most important operating parameter of this detector technology. We focused on simulating different gas compounds $Ar:CO_2$, $Ar:CH_4$, and $Ar:CF_4$ in which we varied the proportion of Argon in each mixture from 50\% to 100\%.
\begin{figure}
	\includegraphics[width=.85\columnwidth]{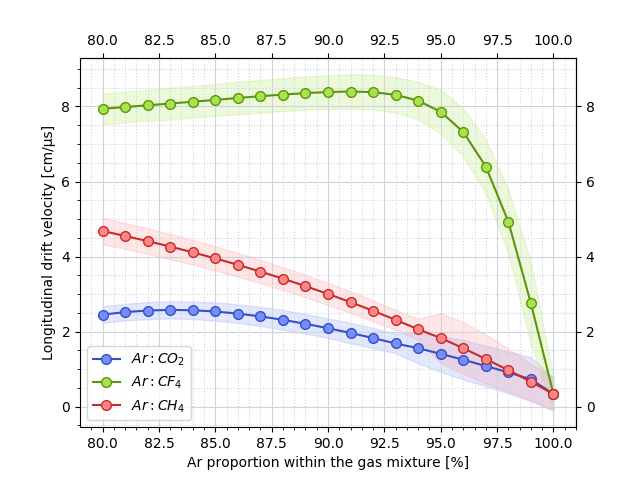}
	\centering
	\caption{Longitudinal drift velocity for distinct gas mixtures and different $Ar$ proportions (between 50\%  and 100\%). Simulation on \SI{5000}{} events with 10 primary electrons per \si{\milli\meter}, and beam angle of \SI{20}{\deg}. Bands represent standard deviation. Cross section tables: $Ar$ from Biagi \cite{biagi2019lxcat}, $CO_2$ from Phelps\cite{phelps2019lxcat}, $CH_4$ from Hayashi \cite{hayashi2019lxcat} and $CF_4$ from Bordage \cite{ bordage2019lxcat}.}
	\label{fig:sim_ratio_velocity}
\end{figure}

\begin{figure}
	\includegraphics[width=.85\columnwidth]{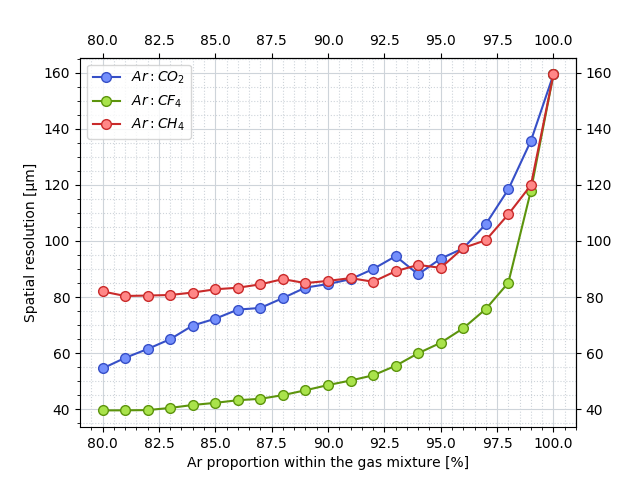}
	\centering
	\caption{Spatial resolution for distinct gas mixtures and different proportions of $Ar$ between 50\%  and 100\%. Simulation on \SI{3000}{} events with \SI[per-mode=symbol]{10}{\per\milli\meter} primary electrons, and beam angle of \SI{20}{\deg}. Cross section tables: $Ar$ from Biagi \cite{biagi2019lxcat}, $CO_2$ from Phelps\cite{phelps2019lxcat}, $CH_4$ from Hayashi \cite{hayashi2019lxcat} and $CF_4$ from Bordage \cite{ bordage2019lxcat}.}
	\label{fig:sim_ratio_resolution}
\end{figure}

The results obtained showed us that the chosen gas mixture has a substantial impact both on longitudinal drift velocity and spatial resolution. 
The $Ar:CO_2$ gas mixture in 93:7 proportion used in the ATLAS experiment represents a good trade-off - lower drift velocity (see fig. \ref{fig:sim_ratio_velocity}) to obtain a great spatial resolution (see fig. \ref{fig:sim_ratio_resolution}). Although the mixture of $Ar: CH_4$ has better values, we should not neglect that $CH_4$ is a flammable gas even at low concentrations, and does not suit being used in this type of detectors where electrical discharges often occur.

\subsection{Simulation of the number of primary particles}

\begin{figure}
	\includegraphics[width=.85\columnwidth]{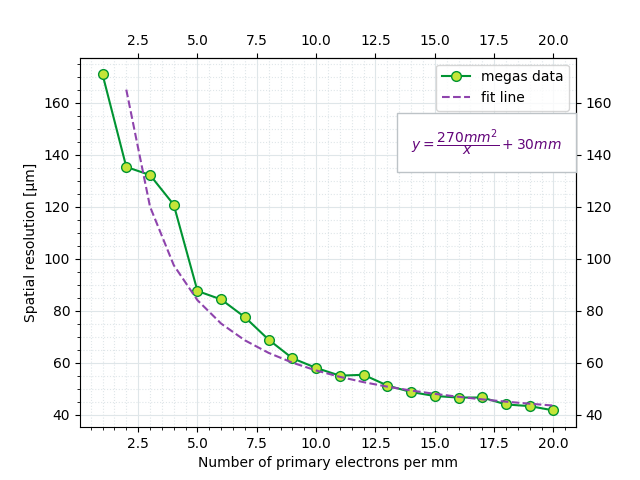}
	\centering
	\caption{Spatial resolution for different number of primary electrons  between 1 and \SI[per-mode=symbol]{20}{\per\milli\meter}. Simulation on \SI{3000}{} events in $Ar:CO_2$ $93:7$ gas mixture at \SI{101.325}{\kilo\pascal}, and beam angle of \SI{20}{\deg}. Cross section tables respectively from Biagi \cite{biagi2019lxcat} and Phelps\cite{phelps2019lxcat}.}
	\label{fig:sim_int_resolution}
\end{figure}

\begin{lstlisting}
megas --angle 20 --events 3000 --interactions $int
\end{lstlisting}
The number of primary electrons generated per \si{\milli\meter} is strictly correlated with the gas mixture and the drift height, and is specific for this detector geometry. Still, for this simulation, we wanted to see how this particular value influences the spatial resolution in this detector (see figure \ref{fig:sim_int_resolution}).

Results shows that the number of primary electrons has a significant impact on the spatial resolution, and they are highly correlated. The minimum number of interactions required to obtain a resolution under \SI{100}{\micro\meter} is five per mm, and 25 for the whole drift region. This distribution can be roughly fitted by the function $y = \frac{270mm^2}{x} + 30mm$.

\section{Conlusions}
\label{sec:conslusions}
We presented \texttt{megas}, a new software tool for modeling Micromegas gas detectors. The application uses a microscopic approach by integrating the \texttt{Betaboltz} library. We simulated several scenarios by modifying the essential parameters like voltage, incident particle angle, gas mixture, and the number of created primary electrons. Each scenario was evaluated by looking at two 
particular properties of the Miromegas detector:  drift velocity along the electrical field and the spatial resolution. 

According to the results presented in this paper, \texttt{megas} can accurately simulate MM detectors yielding results that are comparable with the available experimental data. Simultaneously, our tool can be used to create different simulation scenarios easily, from the number of MM layers to variations in all constructive and operative parameters of the detector.

We demonstrated the functionality of this new simulation tool. With respect to the existing Micromegas detector simulation tools, its significant accomplishments are:
\begin{itemize}
	\item provides a microscopic simulation of electrons in the drift region;
	\item able to reproduce experimental results;
	\item simulation setups easy to customize via input parameters;
	\item provides  improved performances due to faster code execution and support for multi-core execution.
\end{itemize}
\section*{Acknowledgments}
We would like to specially thank Prof. C\u alin Alexa for his valuable suggestions and support, and to thank our IFIN-HH ATLAS group colleagues, for the constructively and friendly environment they create every day. This work was supported by the Romanian Ministry of Research and Innovation thought the research grants \texttt{ATLAS CERN-RO} and \texttt{PN19060104}.

\section*{References}
\bibliography{bibliography}

\end{document}